\documentclass[runningheads, orivec]{llncs}
\usepackage[top=2cm, bottom=2cm, left=3cm, right=3cm]{geometry}
\usepackage{bm}  
\usepackage{amsmath}
\usepackage{amsfonts}

\usepackage{multicol}
\usepackage{color}
\usepackage{float}

\usepackage{braket}
\usepackage[T1]{fontenc}
\usepackage{graphicx}
\usepackage{booktabs}
\usepackage{siunitx}
\usepackage{makecell}
\usepackage{multirow}
\usepackage{array}
\usepackage{enumitem}
\usepackage{longtable}
\usepackage{rotating}
\usepackage{tablefootnote}
\usepackage{graphicx}
\usepackage[hyphens]{url}
\usepackage[colorlinks=true, urlcolor=blue, pdfborder={0 0 0}]{hyperref}
\hypersetup{
  colorlinks   = true, 
  urlcolor     = blue, 
  linkcolor    = blue, 
  citecolor   = blue 
}

\usepackage{caption}
\usepackage{array}
\newcolumntype{C}[1]{>{\centering\arraybackslash}p{#1}}

\begin{document}
\title{Liquidity provision in CLMMs: evidence from transactions data}
\titlerunning{Abbreviated paper title}

\author{Andrey~Urusov\inst{1,3}\thanks{Corresponding author.} \and
Rostislav~Berezovskiy\inst{1}
\and \\
Anatoly~Krestenko\inst{1,2} \and
Andrei~Kornilov\inst{1,4}}

\authorrunning{A. Urusov et al.}
\titlerunning{Liquidity Provision in CLMMs}

\institute{Vega Institute Foundation, Moscow, Russia \\ 
\and 
Moscow Institute of Physics and Technology, Moscow, Russia \\ 
\and 
Chuvash State University, Cheboksary, Russia \\ 
\and
Satbayev Kazakh National Technical University, Almaty, Kazakhstan
}

\maketitle
{
\renewcommand{\thefootnote}{}
\footnotetext{%
  \textit{E‑mail addresses:} tapwi93@gmail.com$^*$ (A.~Urusov),
  rostislavberezovskiy@gmail.com (R.~Berezovskiy),
  acryptokrestenko@gmail.com (A.~Krestenko), 
  kornilov.ag94@gmail.com (A.~Kornilov),
}
}
%
\begin{abstract}
The emergence of Concentrated Liquidity Market Makers (CLMMs) has made liquidity provision on decentralized exchanges an active and risk-sensitive task. However, the standalone profitability of liquidity provision remains unclear for liquidity providers (LPs) who neither hedge their inventory risk nor receive off-pool profits. This paper studies the actual outcomes of LP activity using historical transaction-level data from WETH/USD liquidity pools on the Base chain across the Uniswap, Aerodrome, PancakeSwap and SushiSwap protocols. We propose a methodology for reconstructing LP PnL dynamics from on-chain events and introduce an original metric that captures both the terminal state of LP capital and its path over time. Based on this framework, we estimate the share of successful LPs, classify their behavior and develop a taxonomy of 15 position types as structural components of PnL. We further identify a distinct class of multi-LPs operating across several pools and show that the dominant profitable position configurations are concentrated around the current pool price. The results show that only about one out of six LPs avoids losses in the selected market segment, raising an open question about the true economic motives of LP participation. Evidence also suggests that successful LPs often close positions before the full range is traversed, making observed behavior closer to profit-target-based strategies.

\end{abstract}

\section{Introduction}

\begin{figure*}[!t]
    \centering
    \includegraphics[scale=0.3]{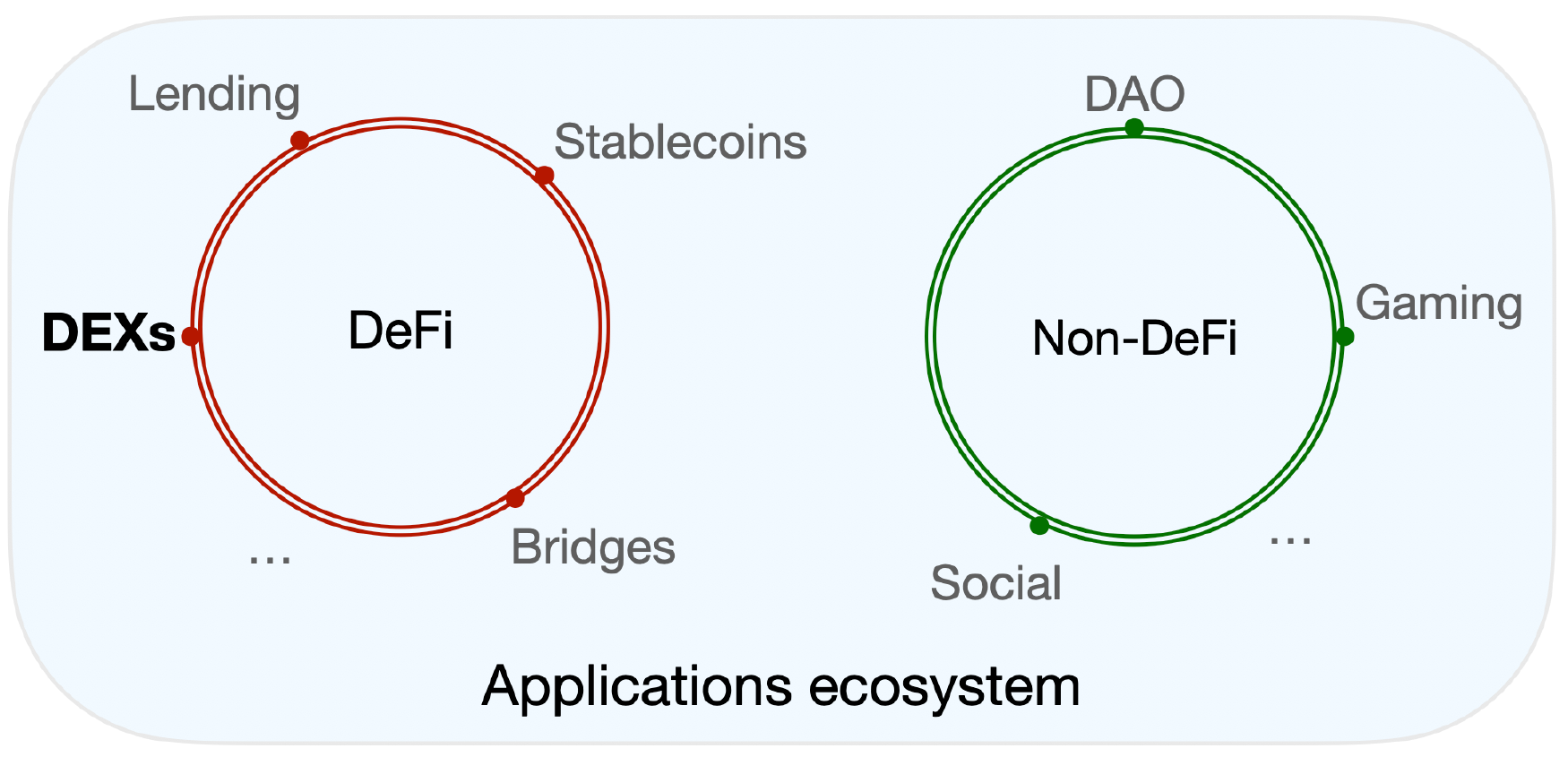}
    \smallskip
    \captionsetup{justification=centering}
    \caption{Position of Decentralized Exchanges in the Blockchain Applications Ecosystem.}
    \label{figure:F1}
\end{figure*}

Blockchain technology has evolved from a narrow infrastructure for peer-to-peer value transfer into a broader ecosystem of decentralized computation, settlement and digital asset exchange. Within this ecosystem, Decentralized Finance (DeFi) has emerged as one of the main application domains, encompassing decentralized exchanges, lending protocols, stablecoins, derivatives and related services \cite{ref_01,ref_02,ref_03}. Among these components, decentralized exchanges (DEXs) occupy a central position as the core venue for on-chain trading, liquidity accumulation and price discovery. As illustrated in Figure \ref{figure:F1}, DEXs should therefore be viewed as a specific class of DeFi applications built on top of a common blockchain infrastructure.

The development of concentrated liquidity market makers (CLMMs) transformed liquidity provision from a largely passive activity into an active management problem. By allowing liquidity providers (LPs) to allocate capital within selected price ranges rather than across the entire price domain, CLMMs significantly improved capital efficiency, but at the cost of increased strategic complexity \cite{ref_04}. In such systems, LPs have to manage price ranges, respond to market movements and frequently relocate capital as prices evolve to remain competitive and continue to earn rewards. Recent research shows that this design makes liquidity provision substantially more demanding, with realized LP outcomes depending on volatility, rebalancing intensity and the balance between fee income and inventory-related losses \cite{ref_05,ref_06}. However, the economic viability of the standalone liquidity provision remains unclear. In particular, previous evidence suggests that active liquidity reallocation may accelerate capital depletion due to adverse capital revaluation and impermanent loss, as shown in study \cite{ref_07} for $\tau$-reset strategies. At the same time, much less is known about how LPs actually behave in historical data: whether profitable users operate within a single pool or across multiple venues, whether they concentrate liquidity around the current pool price and whether position closure is mainly triggered by risk events or by earlier profit-taking decisions.

This paper investigates the profitability of real LP behavior using historical transaction-level data from WETH/USD pools on the Base chain across the Uniswap, Aerodrome, PancakeSwap and SushiSwap protocols. Under the assumption that all gains and losses are generated strictly within the pool, without external hedging or off-pool income, we estimate the share of successful LPs, classify LP behavior and introduce a taxonomy of 15 LP position types. In addition, we identify a distinct class of multi-LPs who allocate liquidity across several pools and analyze the dominant structural features of profitable positions. This makes it possible not only to assess the fraction of successful LPs but also to clarify whether the observed LP behavior is more consistent with classical boundary-based reset rules or with more flexible strategies involving early position closure. Our results indicate that only about 1 out of 6 LPs avoid losses in the selected market segment, suggesting that LP participation is likely driven by broader motives than fee income alone. We further show that the dominant position configuration is liquidity placement around the current pool price, typically with the price located near the midpoint of the selected range, while successful LPs often close positions well before the price crosses the price range.

The remainder of the paper is organized as follows. Section \ref{section:background} describes the data and the empirical framework. Section \ref{section:relwk} reviews the related literature. Section \ref{section:problem_solution} reports the empirical results at the levels of LP and positions. Section \ref{section:Conclusions} discusses the broader implications of the findings and concludes.

\section{Background}
\label{section:background}
This section describes the methodology used to reconstruct and analyze LP PnL from historical transaction-level data. For each liquidity provider, transactions are first partitioned by unique liquidity ranges, after which historical positions are reconstructed using a dynamic balance approach. These positions were then aggregated across all active ranges to obtain the cumulative PnL of the LP during the sample period. The complete set of transactions included in this reconstruction is called the LP strategy. LP performance is assessed by the terminal PnL over the active historical period and by a custom strategy-performance metric denoted as the win-score.

\subsection{Construction of LP PnL}
\label{section:LPPNL}

\subsubsection{Input data.} To reconstruct LP PnL, we use transaction-level data for the three main liquidity operations --- \textit{mint}, \textit{burn} and \textit{collect} --- along with \textit{swap} transactions. Swap events are used to identify the pool price at the moment of each liquidity operation. The dataset is structured at the pool level and the analysis covers a total of \(N\) pools.

\subsubsection{Unique LPs.} Within each pool, we identify unique liquidity providers. We consider only those LPs that have at least one \textit{mint} and one \textit{burn} transaction. This is a necessary, though not sufficient, condition for the reconstruction of LP PnL. This filtering excludes LPs for which PnL cannot be reliably reconstructed within the selected historical period, either because the observation window truncates their activity or because of specific features of their strategy. After this step, each pool \(c\) contains \(m_c\) unique LPs, where \(c \in [1, N]\). The \(i\)-th liquidity provider in pool \(c\) is denoted by \(LP_i^c\), where \(i \in [1, m_c]\). Hereafter, we assume that all objects are specified for a particular liquidity provider \(LP_i^c\) and omit this index completely to avoid notation overcomplication. 

\subsubsection{LP-liquidity ranges.} Each \(LP_i^c\) is associated with a finite set of liquidity ranges in which the provider placed liquidity during the historical sample period. We denote the set of unique ranges of \(LP_i^c\) by
\[
\left\{r_1,\dots,r_{k}\right\},
\]
where \(k\) is the number of unique liquidity ranges used by the provider during the observation period. Each range \(r_j\), where \(j \in [1,k]\), is characterized by its lower and upper boundaries, which may be represented in ticks as \([t_{a_j}, t_{b_j}]\) or equivalently in pool prices as \([p_{a_j}, p_{b_j}]\).

\subsubsection{Matching \textit{burn} and \textit{collect} transactions.} When an LP removes liquidity, the corresponding token amounts and accrued pool rewards are realized through a two-step mechanism at the transaction level. A \textit{burn} operation changes the size of the current position, while a \textit{collect} operation represents the actual withdrawal of tokens available to the LP, including both the position tokens and the accrued fees. In the subsequent analysis, we reconstruct the dynamic balance of an LP position within a given range using the liquidity value (\textit{amount} in the data), which is available for \textit{burn} transactions but absent for \textit{collect} transactions. Therefore, a \textit{collect} transaction cannot be directly matched with a preceding \textit{mint} operation. To resolve this, we first match \textit{burn} and \textit{collect} transactions across all ranges for each LP. We refer to the resulting matched liquidity-removal and payout event as \textit{burn+}.

\subsubsection{Dynamic balance and transaction accounting.} Within each unique range \(r_j\) of the provider \(LP_i^c\), we consider the set of \textit{mint} and matched \textit{burn+} operations, denoted by
\[
\left\{tx_1,\dots,tx_{n_j}\right\},
\]
where \(n_j\) is the number of transactions of provider \(LP_i^c\) in the range \(r_j\) during the historical sample period.
The dynamic balance reconstruction procedure for the range \(r_j\) is as follows:
\begin{enumerate}
    \item The transactions in $\left\{tx_1,\dots,tx_{n_j}\right\}$ are ordered chronologically.
    
    \item Liquidity associated with \textit{mint} operations is recorded with a positive sign, while liquidity associated with \textit{burn+} operations is recorded with a negative sign.
    
    \item Taking the temporal structure into account, we exclude \textit{burn+} operations whose liquidity exceeds the cumulative liquidity introduced by \textit{mint} operations up to that moment. Such events are interpreted as referring to liquidity that was deposited before the beginning of the observation period and therefore cannot be consistently reconstructed within the sample.
    
    \item We then select the first \textit{mint} operation in the ordered set $\left\{tx_1,\dots,tx_{n_j}\right\}$. Its liquidity is subsequently reduced by following \textit{burn+} operations until the position is fully closed.
    
    \begin{enumerate}
        \item Given the pool price at the opening of the first position, denoted by \(P_{M_1}\), and the corresponding token amounts \((x_{M_1}, y_{M_1})\), we compute the capital of \(LP_i^c\) deposited in this position, expressed in numeraire \(y\), as
        \[
        W_{j_{M_1}}=y_{M_1}+x_{M_1}P_{M_1}.
        \]
        This capital corresponds to the amount of liquidity \(L_{j_{M_1}}\).
        
        \item Each matched \textit{burn+} operation reduces \(L_{j_{M_1}}\). For every such operation, indexed by \(s\), with token amounts \((x_{j_s}, y_{j_s})\) at the pool price \(P_{j_s}\), we define the realized capital in numeraire \(y\) as
        \[
        B_{j_s}=y_{j_s}+x_{j_s}P_{j_s}.
        \]
        
        \item Once the initial liquidity \(L_{j_{M_1}}\) is fully exhausted, the position is considered closed. Its closing capital amounts to
        \[
        Q_{j_{M_1}}=\sum_s B_{j_s}.
        \]
        Here, the number of contributing indices \(s \geq 1\). If the final liquidity of the \textit{burn+} transaction exceeds the remaining liquidity \textit{mint} needed to close the current position, its capital is split proportionally between the final closing of \(L_{j_{M_1}}\) and the next position \(L_{j_{M_2}}\), if such a position exists.
        
        \item The realized PnL of the closed position is then given by
        \[
        PnL_{j_{M_1}}=Q_{j_{M_1}}-W_{j_{M_1}}.
        \]
    \end{enumerate}
    
    \item Step 4 is repeated for the next \textit{mint} operation until all available \textit{burn+} operations are exhausted or the last \textit{mint} in the current range has been fully closed.
\end{enumerate}

\begin{figure*}[!t]
    \centering
    \includegraphics[scale=0.5]{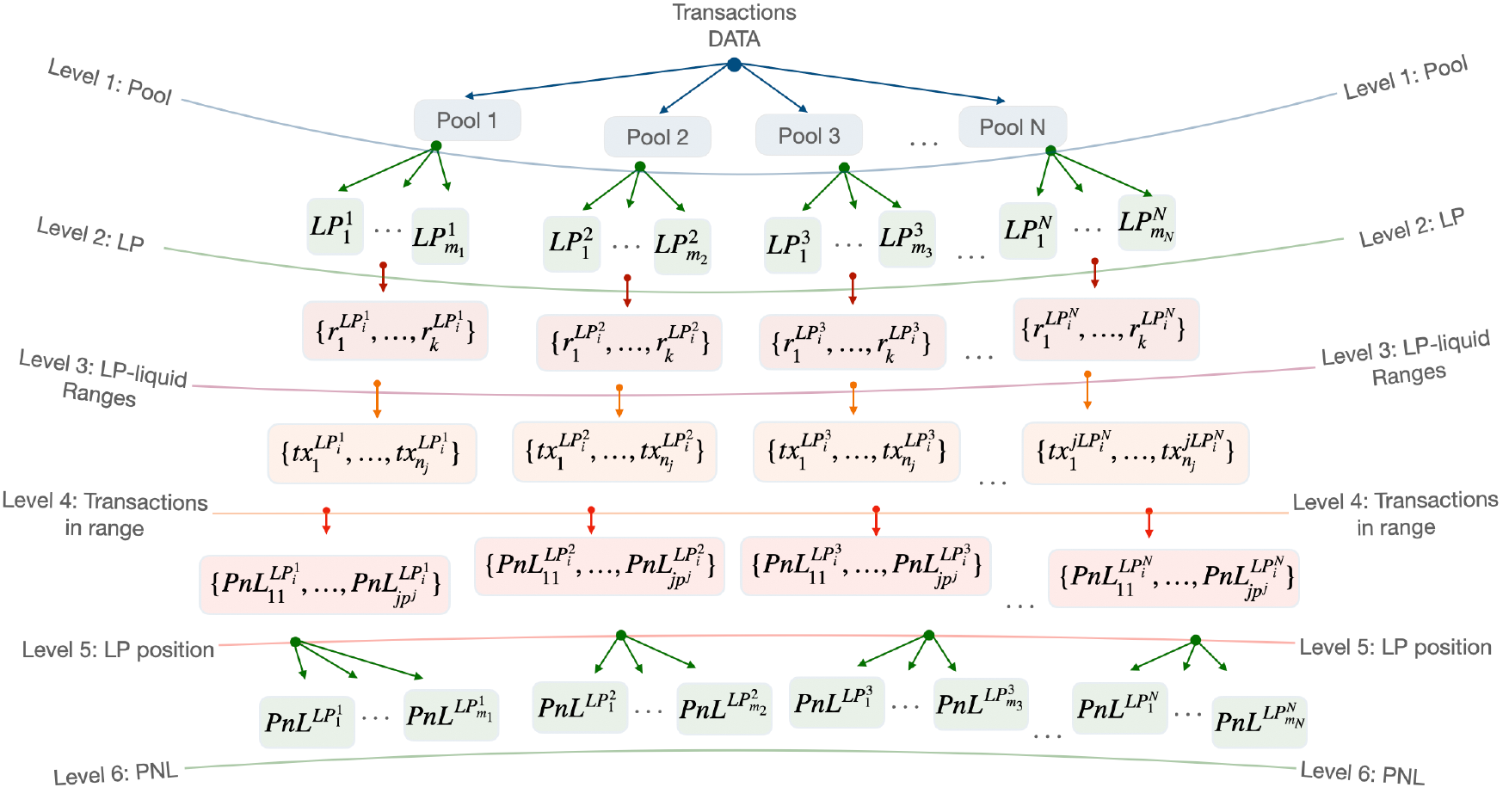}
    \smallskip
    \captionsetup{justification=centering}
    \caption{LP PnL Framework.}
    \label{figure:F2}
\end{figure*}

The same procedure is then applied to all ranges in $\left\{r_1,\dots,r_{k}\right\}$. The final result for the provider \(LP_i^c\) during the historical sample period is defined as

\begin{equation}
PnL
=
\sum_{j=1}^{k}
\sum_{\alpha=1}^{p^{\,j}}
PnL_{j_{M_\alpha}},
\tag{1}
\end{equation}
where \(p^{\,j}\) denotes the number of closed positions in the range \(r_j\). The reconstruction procedure described above is illustrated schematically in Figure \ref{figure:F2}.

\subsection{Win-score metric $\omega$}
\label{section:WINSCORE}

The terminal value of PnL over the historical sample period may be insufficient as a standalone measure of the success of \(LP_i^c\) for at least two reasons. For example, an LP may have PnL>0 while remaining in a loss-making state for most of the observation period, so that the final positive result is largely incidental. Conversely, an LP may have PnL<0 while remaining profitable during most of the period, with the final loss driven primarily by adverse factors near the end of the strategy horizon.

To identify \emph{consistently successful} LPs, we introduce the \emph{win-score} metric \(\omega\), which captures the temporal structure of LP PnL and is described in \ref{app:App11}. By construction,
\[
0 \le \omega \le 1.
\]

If \(\omega>0.5\), \(LP_i^c\) spends a larger share of the observation period in a positive cumulative PnL state; conversely, if \(\omega<0.5\), negative cumulative performance dominates. In the limiting case where all reconstructed positions are win-positions, the metric is equal to \(1\); if all reconstructed positions are loss-positions, it is equal to \(0\).

Together with the terminal value of LP PnL, the proposed metric provides a convenient basis for identifying \emph{consistently successful} LPs by accounting not only for the final result but also for its temporal structure.

\begin{figure*}[]
    \centering
    \includegraphics[scale=0.5]{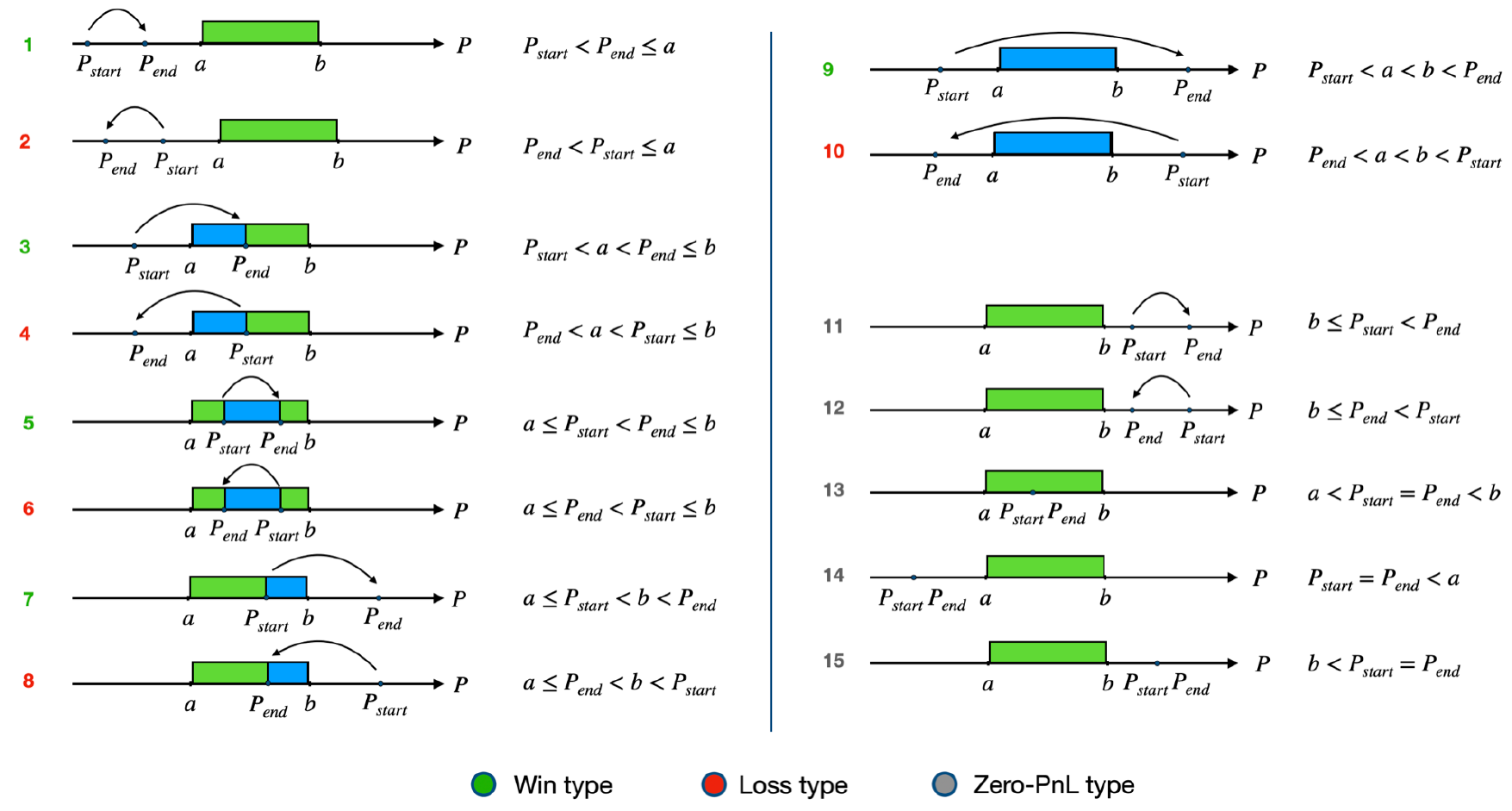}
    \smallskip
    \captionsetup{justification=centering}
    \caption{A schematic grouping of position types.}
    \label{figure:F3}
\end{figure*}

\subsection{Position types}
\label{section:POST}

After classifying LPs by the results of their strategies, we turn to the analysis of individual positions as structural components of the final LP PnL. We propose a classification of positions into 15 types, reflecting the location of the pool price at the time of position opening \(P_{\mathrm{start}}\) and the capital-weighted closing price of the position 
$
P_{\mathrm{end}},
$
relative to the liquidity range boundaries \([a,b]\).

Figure \ref{figure:F3} presents a schematic grouping of position types into three broad categories:
\begin{itemize}
    \item \emph{win} positions, which are profitable for the LP;
    \item \emph{loss} positions, which are unprofitable for the LP;
    \item \emph{PnL-neutral positions}, for which the market value of the capital allocated as liquidity remains unchanged.
\end{itemize}

To characterize the main position types, we introduce the following original metrics, which describe the dynamics of the position relative to the liquidity range and its boundaries.

\textbf{Delta metric.} This metric captures the share of the LP liquidity range that is traversed during the lifetime of the position. It is applicable to Types 3--8 and is defined as
\[
\Delta_{\mathrm{start}}
=
\min\!\left(
\max\!\left(
\frac{P_{\mathrm{start}}-a}{b-a},\,0
\right),\,1
\right),
\qquad
\Delta_{\mathrm{end}}
=
\min\!\left(
\max\!\left(
\frac{P_{\mathrm{end}}-a}{b-a},\,0
\right),\,1
\right),
\]
\begin{equation}
\Delta = \Delta_{\mathrm{end}} - \Delta_{\mathrm{start}}.
\tag{2}
\end{equation}

\textbf{Metrics of the price position relative to the range boundaries at opening and closing.} These metrics are applicable to Types 1--10 and are defined as
\begin{equation}
m_a = \frac{P_{\mathrm{start}}}{a} - 1,
\qquad
m_b = \frac{P_{\mathrm{start}}}{b} - 1,
\tag{3.1}
\end{equation}
\begin{equation}
d_a = \frac{P_{\mathrm{end}}}{a} - 1,
\qquad
d_b = \frac{P_{\mathrm{end}}}{b} - 1.
\tag{3.2}
\end{equation}




\section{Related Work}
\label{section:relwk}

The behavior of liquidity providers on CLMM exchanges is being studied thoroughly in the literature. This topic is of utmost importance not only from the theoretical point of view but also for practice, since active liquidity provision is a widely spread investment strategy in cryptocurrency markets. The concept of CLMMs is described in detail in the works \cite{ref_04}, \cite{ref_08}, \cite{ref_09}, and in our previous paper on this topic  \cite{ref_07}. The paper \cite{ref_04} introduces the very concept of CLMMs, the works \cite{ref_08} and \cite{ref_09} explore in detail the liquidity calculations and the expressions for the capital dynamics and the amount of tokens stored in the price ranges. 

Extensive research exists on active liquidity management strategies in CLMMs. The papers \cite{ref_11}, \cite{ref_12} work with $\tau$-reset strategies, active liquidity management, and accumulated fees in the dynamic setting. In paper \cite{ref_122}, the authors propose a method for approximating historical liquidity for modeling strategies. Moreover, in paper \cite{ref_13} the authors propose a deep learning approach to achieve an efficient active liquidity reallocation strategy. 

The other side of the liquidity provision activity is the risk incurred by the LP. The main risk is impermanent loss studied in \cite{ref_14}. It evolved even further into the discussions about LVR (loss versus rebalancing) proposed in \cite{ref_15} that are extended in the papers \cite{ref_16} and \cite{ref_17}.

The papers mentioned above dive deep into concentrated liquidity and study the risks of liquidity provision, but the general literature lacks detailed studies about the liquidity microstructure in CLMM markets in a dynamic setting. The behavior of multiple liquidity providers logically evolves into game theoretic models to study the interactions between the LPs and the corresponding equilibrium liquidity profiles. This approach is gaining increasing attention among researchers, particularly an instructive model of a game theoretic equilibrium with multiple price ranges may be found in \cite{ref_18}.

One of the important applications of this work is to verify and support the theoretical results on the dynamic liquidity profile in a CLMM as a result of collective LP activity. Such ideas are presented in the work \cite{ref_10} in which the authors use the mean-field approach to describe the resulting liquidity in the pool and the equilibrium strategy of each liquidity provider. 

\section{Results}
\label{section:problem_solution}

This section presents the results of the LP classification by degree of success, together with an analysis of their historical positions based on the methodology described in Section \ref{section:background}. Jupyter notebooks with code and raw data are available on GitHub \cite{BBcs}.

\subsection{Input data and transaction analysis}
We analyze transaction-level data from the Base network for WETH/USD pools across four decentralized exchanges: Uniswap, Aerodrome, PancakeSwap and SushiSwap. The Base network is chosen for two reasons: first, it demonstrates high TVL competitive with the Ethereum network for popular DEX markets; second, it has relatively low fees, making it possible to expect various active liquidity management strategies applied in practice. The historical sample covers the period from September~2024 to July~2025. Table \ref{tab:input_data} reports the main characteristics of the initial input data. 

\begin{table*}
\centering
\caption{Summary Statistics of the Initial Transaction Data}
\label{tab:input_data}
\small
\setlength{\tabcolsep}{6pt}
\renewcommand{\arraystretch}{1.15}
\begin{tabular}{lccccc}
\toprule
\multirow{3}{*}{Event} 
& \multicolumn{4}{c}{Base WETH/USDC} 
& \multirow{3}{*}{Total} \\
\cmidrule(lr){2-5}
& \makecell{Uniswap V3\\0.05\%\footnotemark[1]}
& \makecell{Aerodrome\\SlipStream 0.05\%\footnotemark[2]}
& \makecell{PancakeSwap V3\\0.01\%\footnotemark[3]}
& \makecell{SushiSwap V3\\0.01\%\footnotemark[4]}
& \\
\midrule
Swap     & \num{4473460} & \num{3584871} & \num{9070911} & \num{4922958} & \num{22052200} \\
Mint     & \num{484032}  & \num{1465004} & \num{24456}   & \num{39917}   & \num{2013409}  \\
Burn     & \num{786921}  & \num{6090188} & \num{27728}   & \num{38783}   & \num{6943620}  \\
Collect  & \num{754211}  & \num{6074987} & \num{27392}   & \num{39948}   & \num{6896538}  \\
Other    & \num{583}     & \num{46}      & \num{1530}    & \num{14935}   & \num{17094}    \\
\midrule
transactions & \num{6499207} & \num{17215096} & \num{9152017} & \num{5056541} & \num{37922861} \\
LP count & \num{21863}   & \num{15042}    & \num{7023}    & \num{625}     & \num{44553}    \\
\bottomrule
\end{tabular}
\end{table*}

\footnotetext[1]{Contract 0xd0b53d9277642d899df5c87a3966a349a798f224}
\footnotetext[2]{Contract 0xb2cc224c1c9fee385f8ad6a55b4d94e92359dc59}
\footnotetext[3]{Contract 0x72ab388e2e2f6facef59e3c3fa2c4e29011c2d38}
\footnotetext[4]{Contract 0x482fe995c4a52bc79271ab29a53591363ee30a89}

\textit{Swap} transactions were used solely to identify the contemporaneous pool price at the time of each liquidity operation. All subsequent stages of the analysis relied only on LP-related operations, namely \textit{mint}, \textit{burn} and \textit{collect}. The data suggest that pools with lower fee levels are more attractive to traders, whereas pools with higher fee levels appear to be more attractive to LPs, which is consistent with economic intuition.

Zero-value transactions were excluded from the dataset. It is worth noting that such operations account for approximately 50\% of the initial historical liquidity-related transactions. After this filtering step, the resulting sample can be treated as the cleaned dataset used for the empirical analysis; its main characteristics are reported in Table \ref{tab:clean_data}.

\begin{table*}[]
\centering
\caption{Main Characteristics of the Data After Technical Filtering}
\label{tab:clean_data}
\small
\setlength{\tabcolsep}{6pt}
\renewcommand{\arraystretch}{1.15}
\begin{tabular}{lccccc}
\toprule
\multirow{3}{*}{Event} 
& \multicolumn{4}{c}{Base WETH/USDC} 
& \multirow{3}{*}{Total} \\
\cmidrule(lr){2-5}
& \makecell{Uniswap V3\\0.05\%}
& \makecell{Aerodrome\\SlipStream 0.05\%}
& \makecell{PancakeSwap V3\\0.01\%}
& \makecell{SushiSwap V3\\0.01\%}
& \\
\midrule
Mint     & 484 026   & 1 464 996 & 24 456 & 39 917 & 2 013 395 \\
Burn     & 536 265   & 2 266 316 & 18 478 & 37 398 & 2 858 457 \\
Collect  & 724 737   & 2 357 378 & 26 208 & 38 970 & 3 147 293 \\
\midrule
trans. & 1 745 028 & 6 088 690 & 69 142 & 116 285 & 8 019 145 \\
\bottomrule
\end{tabular}
\end{table*}

Applying the algorithm described in Section \ref{section:LPPNL}, approximately 65\% of the non-zero transaction data are retained at the dynamic-balance stage. The main statistics of the transactions used are reported in Table \ref{tab:used_data}.

\begin{table*}[]
\centering
\caption{Summary Statistics of the Used Transaction Data}
\label{tab:used_data}
\small
\setlength{\tabcolsep}{6pt}
\renewcommand{\arraystretch}{1.15}
\begin{tabular}{lccccc}
\toprule
\multirow{3}{*}{Event} 
& \multicolumn{4}{c}{Base WETH/USDC} 
& \multirow{3}{*}{Total} \\
\cmidrule(lr){2-5}
& \makecell{Uniswap V3\\0.05\%}
& \makecell{Aerodrome\\SlipStream 0.05\%}
& \makecell{PancakeSwap V3\\0.01\%}
& \makecell{SushiSwap V3\\0.01\%}
& \\
\midrule
Mint     & 338 760   & 1 359 990 & 5 740  & 23 869 & 1 728 359 \\
Burn     & 346 738   & 1 382 050 & 5 268  & 23 937 & 1 757 993 \\
Collect  & 346 738   & 1 382 050 & 5 268  & 23 937 & 1 757 993 \\
\midrule
trans. & 1 032 236 (19.7\%) & 4 124 090 (78.6\%) & 16 276 (0.3\%) & 71 743 (1.4\%) & 5 244 345\\
\bottomrule
\end{tabular}
\end{table*}

The identical values reported for the \textit{burn} and \textit{collect} types result from the matching of these operations with the composite \textit{burn+} event during the reconstruction procedure. It should also be noted that, under a comparable fee level, the Aerodrome pool appears to be more attractive to LPs than Uniswap in terms of liquidity provision, accounting for about 78.6\% of all retained liquidity-related transactions.

\subsection{LP classification}

The analysis of LP success shows that, despite the strong dominance of Aerodrome in terms of liquidity-related operations, the distribution of users across the analyzed pools follows a different pattern. In particular, Uniswap appears to be more attractive to users overall, accounting for approximately 48.3\% of all unique users in the sample. This suggests that, all else being equal, users --- possibly including newer participants without clearly defined liquidity-management strategies --- tend to prefer the more established Uniswap platform. At the same time, Aerodrome shows a pronounced advantage in the number of liquidity management operations, pointing to a substantially higher intensity of active liquidity management among its LPs.

\begin{table*}[]
\centering
\caption{Main LP Characteristics}
\label{tab:lp_characteristics}
\small
\setlength{\tabcolsep}{6pt}
\renewcommand{\arraystretch}{1.15}
\begin{tabular}{lccccc}
\toprule
\multirow{3}{*}{LP metrics}
& \multicolumn{4}{c}{Base WETH/USDC}
& \multirow{3}{*}{Total} \\
\cmidrule(lr){2-5}
& \makecell{Uniswap V3\\0.05\%}
& \makecell{Aerodrome\\SlipStream 0.05\%}
& \makecell{PancakeSwap\\V3 0.01\%}
& \makecell{SushiSwap\\V3 0.01\%}
& \\
\midrule
count
& 15 841 (48.3\%)
& 11 531 (35.1\%)
& 4 968 (15.1\%)
& 476 (1.5\%)
& 32 816 (100\%) \\
\midrule
$PnL>0$
& 2 629 (16.6\%)
& 1 846 (16.0\%)
& 624 (12.6\%)
& 56 (11.8\%)
& 5 155 (15.7\%) \\
\midrule
$PnL>0$ \& $\mathrm{\omega}>0.5$
& 2 541 (16.0\%)
& 1 757 (15.2\%)
& 613 (12.3\%)
& 51 (10.7\%)
& 4 962 (15.1\%) \\
\bottomrule
\end{tabular}
\end{table*}

As shown in Table \ref{tab:lp_characteristics}, the analyzed pools exhibit the following pattern: on average, only about one out of six LPs achieves a positive outcome from liquidity provision, while only about one out of seven can be classified as consistently successful, with both a non-negative terminal PnL and \(\mathrm{\omega}>0.5\). The remaining users incur losses. In the pools that appear more attractive to LPs, namely Uniswap and Aerodrome, this ratio is broadly consistent with the overall average, whereas in the less liquid pools it deteriorates to approximately one in nine.

The analysis of LP behavior at the network level also reveals specific categories of users who allocated liquidity across multiple pools during the historical period. Table \ref{tab:lp_nature} presents the decomposition of unique LPs across the analyzed pools into the classes of mono-LPs and multi-LPs.

\begin{table*}[]
\centering
\caption{Mono- and Multi-Pool Composition of Unique LPs}
\label{tab:lp_nature}
\small
\setlength{\tabcolsep}{6pt}
\renewcommand{\arraystretch}{1.15}
\begin{tabular}{llccccc}
\toprule
\multirow{3}{*}{\makecell{LP\\category}} 
& \multirow{3}{*}{\makecell{Type}}
& \multicolumn{4}{c}{Base WETH/USDC}
& \multirow{3}{*}{Total} \\
\cmidrule(lr){3-6}
& 
& \makecell{Uniswap V3\\0.05\%}
& \makecell{Aerodrome\\SlipStream 0.05\%}
& \makecell{PancakeSwap V3\\0.01\%}
& \makecell{SushiSwap V3\\0.01\%}
& \\
\midrule
\multicolumn{2}{l}{pool's LP} 
& 15 841 & 11 531 & 4 968 & 476 & 32 816 \\
\midrule
\multirow{2}{*}{\makecell[l]{uniq. on\\network}}
& mono  
& 13 250 & 9 721 & 3 508 & 293 & \multirow{2}{*}{29 695} \\
\cline{2-6}
& multi 
& \multicolumn{4}{c}{2 923} & \\
\bottomrule
\end{tabular}
\end{table*}

\begin{figure*}[]
    \centering
    \includegraphics[scale=0.32]{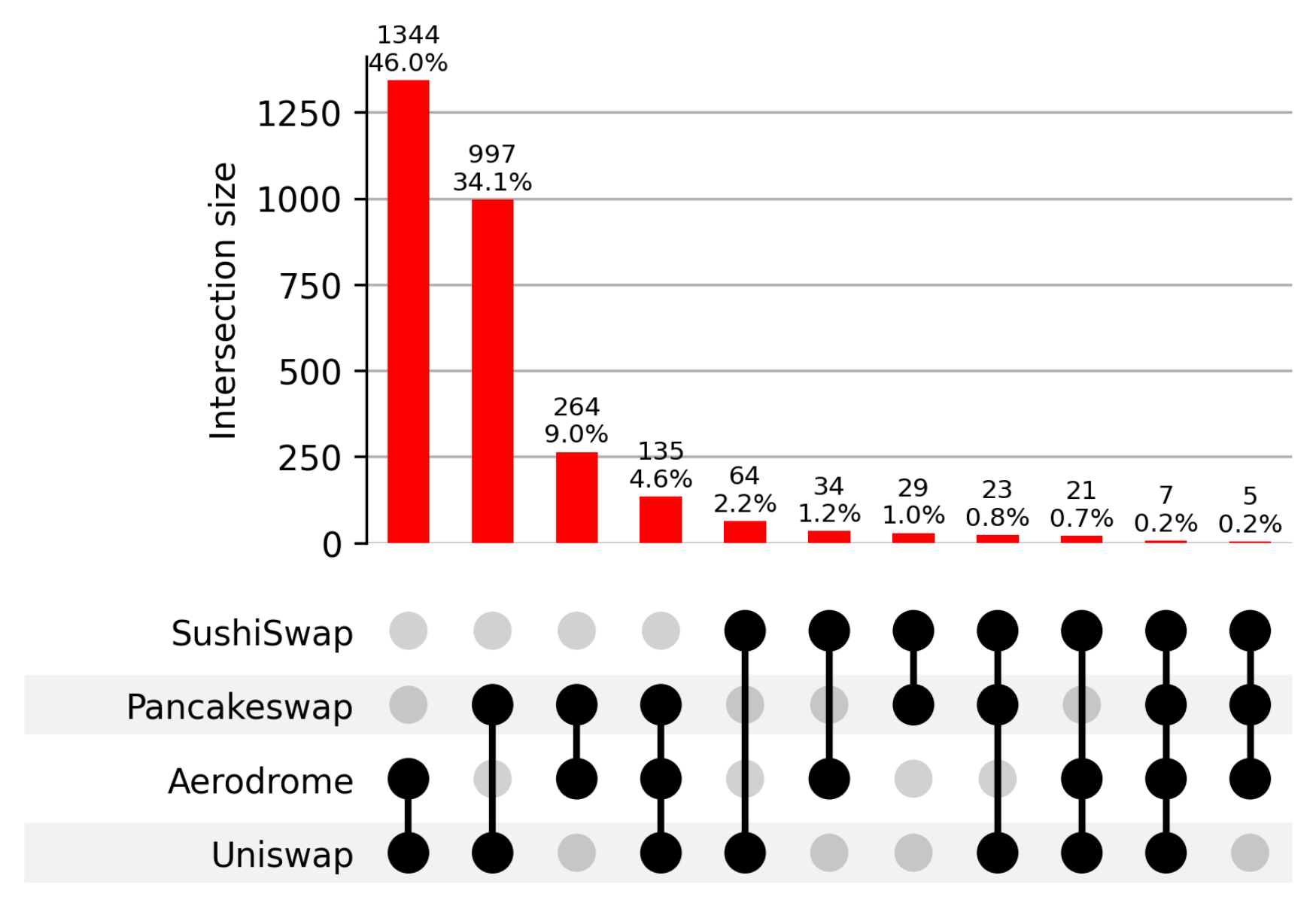}
    \smallskip
    \captionsetup{justification=centering}
    \caption{Cross-Pool Distribution of Multi-LPs in the Analyzed WETH/USD Pools.}
    \label{figure:F5}
\end{figure*}

The results show that approximately 10\% of all unique LPs retained in the historical sample allocated liquidity in more than one of the analyzed pools. Figure \ref{figure:F5} illustrates the cross-pool migration of users across the considered pools.

\begin{table*}[!b]
\centering
\caption{Cross-Pool Participation and Performance of Multi-LPs}
\label{tab:multi_lp_results}
\small
\setlength{\tabcolsep}{5pt}
\renewcommand{\arraystretch}{1.15}
\begin{tabular}{p{2.2cm} p{3.1cm} ccccc}
\toprule
\multirow{2}{*}{Group by} & \multirow{2}{*}{LP metrics}
& \multicolumn{4}{c}{Number of pools}
& \multirow{2}{*}{Total} \\
\cmidrule(lr){3-6}
& & one & two & three & four & \\
\midrule
participated
& count
& --
& 2 732 (93.5\%)
& 184 (6.3\%)
& 7 (0.2\%)
& 2 923 \\
\midrule
\multirow{2}{=}{profitable}
& $PnL>0$
& 891 (30.5\%)
& 114 (3.9\%)
& 1 (0.03\%)
& --
& -- \\
& $PnL>0$ \& $\omega>0.5$
& 863 (29.5\%)
& 105 (3.6\%)
& 1 (0.03\%)
& --
& -- \\
\bottomrule
\end{tabular}
\end{table*}

It can be seen that cross-pool activity is concentrated primarily in combinations involving the more liquid Aerodrome and Uniswap pools. At the same time, the share of multi-LPs present in all four pools is only about 0.2\%, while in the vast majority of cases (93.5\%) cross-pool activity is limited to two pools. The main results for multi-LPs are reported in Table \ref{tab:multi_lp_results}.

A consistently positive outcome in at least one of the cross-pools is observed for 29.5\% of multi-LPs. At the same time, the increase in cross-pool connectivity is associated with a sharp decline in performance: only 3.6\% of multi-LPs remain consistently successful in two or more pools, 0.03\% in three or more pools, and none across all pools simultaneously.

A more detailed analysis of multi-LP activity over the considered period does not reveal a substantial advantage of this user category in terms of consistent performance relative to mono-LPs, partly due to the limited empirical base available for such comparisons. At the same time, the identification of the multi-LP class is an important step toward understanding the broader motivation of users, since such cross-pool activity may indicate the presence of more complex liquidity-providing strategies.

\subsection{Classification of Individual Positions}

All reconstructed positions involved in the construction of LP PnL were classified into 15 types according to Section \ref{section:POST}. Table \ref{tab:position_type_distribution} reports the distribution of utilized transactions and of the capital, expressed in the selected numeraire, allocated at position opening across types. The distribution is presented at the aggregate level for all analyzed pools.

\begin{table}[]
\centering
\caption{Distribution of Transactions and Capital Across Position Types}
\label{tab:position_type_distribution}
\small
\setlength{\tabcolsep}{8pt}
\renewcommand{\arraystretch}{1.15}
\begin{tabular}{ccc}
\toprule
\multirow{2}{*}{Position type} & \multicolumn{2}{c}{Share of} \\
\cmidrule(lr){2-3}
& transactions & mint capital \\
\midrule
1  & 0.4\%   & 0.02\%  \\
2  & 1.0\%   & 0.02\%  \\
3  & 0.5\%   & 0.1\%   \\
4  & 2.2\%   & 0.3\%   \\
5  & 45.8\%  & 48.4\%  \\
6  & 45.0\%  & 49.2\%  \\
7  & 2.1\%   & 0.3\%   \\
8  & 0.5\%   & 0.3\%   \\
9  & 0.5\%   & 0.0\%   \\
10 & 0.4\%   & 0.0\%   \\
11 & 0.4\%   & 0.0\%   \\
12 & 0.1\%   & 0.0\%   \\
13 & 1.2\%   & 1.3\%   \\
14 & 0.001\% & 0.001\% \\
15 & 0.001\% & 0.001\% \\
\bottomrule
\end{tabular}
\end{table}

It can be seen that more than 90\% of transactions and allocated capital are concentrated in Types 5--6, i.e., in situations where the current pool price lies within the LP-selected range boundaries both at position opening and at position closure.

We next focus on the position types corresponding to LP win-positions, namely Types 3, 5 and 7. These types are of primary interest because Type 5 represents the dominant in-range configuration, while Types 3 and 7 capture upward price movements into and out of the selected range. Moreover, for these position types, the $\Delta$ metric is particularly informative, as it characterizes the extent of price moves within the range before position closure. The analysis is restricted to consistently successful LPs, defined as those with positive terminal PnL and \(\mathrm{\omega}>0.5\). For each LP and each type of position, we compute the capital-weighted average width of the position range. The resulting metrics are then analyzed across five categories of weighted position-range width expressed in the numeraire (USDC):
\begin{itemize}
    \item \((0,10)\): narrow ranges;
    \item \([10,100)\): medium ranges;
    \item \([100,1000)\): wide ranges;
    \item \([1000,10000)\): ultra-wide ranges corresponding to the level of theoretically active pool liquidity;
    \item \(>10000\): ultra-wide ranges of the ``Uniswap v2-like'' type.
\end{itemize}

The main significant metrics and the corresponding analytical results for the above win-position types are presented below. 
In the tables below, \(rel_a = d_a - m_a\) and \(rel_b = d_b - m_b\) denote the changes in the price position relative to the lower and upper boundaries of the selected range, respectively; \(R_{\mathrm{price}}\) denotes the price dynamics over the lifetime of the position; and \(R_{\mathrm{cap}}\) denotes the position capital return, i.e., the relative change between the capital received at closure and the capital allocated at opening.
The full set of metrics for all win-position types is reported in \ref{app:App22}.

\paragraph{Position Type 3: \(P_{\mathrm{start}}<a<P_{\mathrm{end}}\leq b\).}\mbox{}\\

\textbf{Interpretation.} In this type of position, liquidity is initially placed in a range located to the right of the current pool price \(P_{\mathrm{start}}\), reflecting the LP's expectation of further price growth and of the subsequent entry of the price into the interval \([a,b]\). Owing to favorable market dynamics, the capital-weighted closing price $P_{\mathrm{end}}$ ends up inside the LP liquidity range.

\textbf{Results and implications.} Positions of this type are characterized by a concentration of capital in \emph{medium-width} ranges, while in terms of the number of transactions the dominant categories are \emph{medium} and \emph{wide} ranges. Table \ref{tab:type3_metrics} reports a concise set of metrics for this position type.
\begin{table*}[]
\centering
\caption{Main Metrics for Position Type 3}
\label{tab:type3_metrics}
\small
\setlength{\tabcolsep}{5pt}
\renewcommand{\arraystretch}{1.15}
\begin{tabular}{lccccccccc}
\toprule
\makecell{Range width\\(USDC)} 
& LPs 
& Trans. 
& TVL 
& \makecell{Duration,\\sec} 
& \makecell{$\Delta$, \%\\(USDC)} 
& $m_a$ 
& $rel_a$ 
& $R_{\mathrm{cap}}$ 
& $R_{\mathrm{price}}$ \\
\midrule
$(0,10)$              & 8.9\%  & 7.5\%  & 5.5\%    & 1 807     & 50\% (3)        & -0.1\%  & 0.2\%  & 0.1\%  & 0.2\%  \\
$[10,100)$           & 17.1\% & 50.4\% & 87.6\%   & 76 020    & 31\% (9)        & -0.1\%  & 0.5\%  & 0.4\%  & 0.5\%  \\
$[10^2,10^3)$        & 59.3\% & 33.9\% & 6.5\%    & 722 239   & 39\% (175)      & -0.8\%  & 8.0\%  & 6.1\%  & 8.2\%  \\
$[10^3,10^4)$        & 13.8\% & 7.9\%  & 0.4\%    & 2 063 699 & 28\% (435)      & -0.3\%  & 26.2\% & 17.0\% & 26.3\% \\
$>10^4$              & 0.8\%  & 0.4\%  & 0.001\%  & 14 580    & 0.0005\% (25)   & -0.03\% & 1.4\%  & 1.4\%  & 1.4\%  \\
\midrule
\textbf{Overall\footnotemark[5]}     &        &        &          & \textbf{} & \textbf{35\% (98)} & \textbf{-0.3\%} & \textbf{5.0\%} & \textbf{3.7\%} & \textbf{5.1\%} \\
\bottomrule
\end{tabular}
\end{table*}

\footnotetext[5]{The reported metrics are additionally weighted by the number of transactions of each LP.}

An important finding is that the weighted delta metric is approximately \(35\%\). This suggests that successful LPs typically do not wait until the market price fully traverses their selected range; instead, they tend to close positions earlier, after achieving a target capital return generated, on average, over roughly one third of the range width. At the same time, the lower boundary of the range \(a\) exceeds the current pool price by about \(0.3\%\) on average, while the cumulative price change relative to the lower boundary is close to \(5\%\).

\paragraph{Position Type 5: \(a \leq P_{\mathrm{start}} < P_{\mathrm{end}} \leq b\).}\mbox{}\\

\textbf{Interpretation.} In this type of position, liquidity is placed in a range that already contains the current pool price \(P_{\mathrm{start}}\), reflecting the LP's expectation of further price growth. Due to favorable market dynamics, the capital-weighted closing price \(P_{\mathrm{end}}\) exceeds \(P_{\mathrm{start}}\) while remaining within the LP liquidity range.

\textbf{Results and implications.} Positions of this type are characterized by a concentration of capital in \emph{narrow} and \emph{medium-width} ranges, while in terms of the number of transactions the dominant category is \emph{narrow} ranges. Table \ref{tab:type5_metrics} reports a concise set of metrics for this position type.

\begin{table*}[]
\centering
\caption{Main Metrics for Position Type 5}
\label{tab:type5_metrics}
\small
\setlength{\tabcolsep}{5pt}
\renewcommand{\arraystretch}{1.15}
\begin{tabular}{lccccccccc}
\toprule
\makecell{Range width\\(USDC)} 
& LPs 
& Trans. 
& TVL 
& \makecell{Duration,\\sec} 
& \makecell{$\Delta$, \%\\(USDC)} 
& \makecell{$\Delta_{start}$} 
& $rel_a$ 
& $R_{\mathrm{cap}}$ 
& $R_{\mathrm{price}}$ \\
\midrule
$(0,10)$       & 0.4\%  & 75.0\% & 58.5\% & 70 524    & 8\% (1)         & 49.5\% & 0.02\%        & 0.01\% & 0.0\%  \\
$[10,100)$     & 7.8\%  & 23.0\% & 41.1\% & 15 963    & 9\% (2)         & 43.0\% & 0.1\%         & 0.05\% & 0.1\%  \\
$[10^2,10^3)$  & 67.9\% & 1.4\%  & 0.3\%  & 331 258   & 21\% (83)       & 44.6\% & 3.7\%         & 1.5\%  & 3.4\%  \\
$[10^3,10^4)$  & 15.5\% & 0.6\%  & 0.03\%  & 865 781   & 13\% (201)      & 46.0\% & 20.0\%        & 3.2\%  & 8.0\%  \\
$>10^4$        & 8.3\%  & 0.1\%  & 0.001\%  & 2 108 464 & 0.006\% (276)   & 0.1\%  & $>10^4\%$     & 6.1\%  & 12.7\% \\
\midrule
\textbf{Overall} 
               &        &        &        & 
               & \textbf{8\%} 
               & \textbf{47.8\%} 
               & \textbf{0.2\%\footnotemark[6]} 
               & \textbf{0.1\%} 
               & \textbf{0.1\%} \\
\bottomrule
\end{tabular}
\end{table*}

\footnotetext[6]{Without the last one position type.}

An important feature of this position type is the delta metric. In contrast to Types 3 and 7, it is more moderate here and reflects the full trajectory of the pool price within the position over its lifetime. The results show that successful LPs typically choose the range boundaries \([a,b]\) so that the current pool price at entry is located close to the midpoint of the interval, and the mint-place metric averaging about \(47.8\%\). After reaching a target level of capital return, the position is usually closed after the price has moved, on average, only about \(8\%\) of the range width. The large values of the \(rel_a\) metric are explained by the fact that, for this position type, the lower boundary \(a\) is typically close to zero.

\paragraph{Position Type 7: \(a \leq P_{\mathrm{start}} \leq b < P_{\mathrm{end}}\).}\mbox{}\\

\textbf{Interpretation.} In this type of position, liquidity is placed in a range that contains the current pool price \(P_{\mathrm{start}}\), reflecting the LP's expectation of further price growth. Due to favorable market dynamics, the capital-weighted closing price \(P_{\mathrm{end}}\) exceeds both the opening price \(P_{\mathrm{start}}\) and the upper boundary \(b\) of the LP liquidity range.

\textbf{Results and implications.} Positions of this type are characterized by a concentration of capital in \emph{medium-width} ranges, while in terms of the number of transactions, the dominant category is \emph{wide} ranges. If LPs whose positions fall into Types 3, 5, and 7 are viewed as a separate group, the preferred strategy appears to be liquidity provision in \emph{wide} ranges. Table \ref{tab:type7_metrics} reports a concise set of metrics for this position type.
\begin{table*}[]
\centering
\caption{Main Metrics for Position Type 7}
\label{tab:type7_metrics}
\small
\setlength{\tabcolsep}{5pt}
\renewcommand{\arraystretch}{1.15}
\begin{tabular}{lccccccccc}
\toprule
\makecell{Range width\\(USDC)} 
& LPs 
& Trans. 
& TVL 
& \makecell{Duration,\\sec} 
& \makecell{$\Delta_{start}$} 
& $d_b$ 
& $rel_b$ 
& $R_{\mathrm{cap}}$ 
& $R_{\mathrm{price}}$ \\
\midrule
$(0,10)$       & 4.8\%  & 4.5\%  & 5.9\%  & 886 633   & 58.1\% & 6.5\% & 6.5\%  & 1.84\% & 6.6\%  \\
$[10,100)$     & 17.3\% & 26.2\% & 75.1\% & 204 599   & 62.6\% & 1.8\% & 2.5\%  & 0.61\% & 2.5\%  \\
$[10^2,10^3)$  & 75.7\% & 67.8\% & 18.3\% & 633 072   & 53.9\% & 3.4\% & 8.6\%  & 1.9\%  & 9.3\%  \\
$[10^3,10^4)$  & 2.3\%  & 1.5\%  & 0.8\%  & 1 962 903 & 66.8\% & 3.0\% & 15.9\% & 4.5\%  & 20.1\% \\
$>10^4$        & 0.0\%  & 0.0\%  & 0.0\%  & --        & --     & --    & --     & --     & --     \\
\midrule
\textbf{Overall} 
               &        &        &        & 
               & \textbf{56.6\%} 
               & \textbf{3.15\%} 
               & \textbf{6.99\%} 
               & \textbf{1.57\%} 
               & \textbf{7.59\%} \\
\bottomrule
\end{tabular}
\end{table*}

In contrast to Type 5, the pool price at the moment of position opening is shifted somewhat closer to the upper boundary, with the mint-place metric averaging \(56.6\%\). After the price rises above the upper boundary \(b\), position closure does not occur immediately; instead, the position is typically closed only after the cumulative price change relative to \(b\) reaches approximately \(7\%\). For relatively small values of the \(d_b\) metric, this type of position may be described by a \(\tau\)-reset strategy, whereas for larger values of \(d_b\) it is more consistent with a \(\tau+\eta^{up}\)-reset strategy, as described in \cite{ref_07}.

\subsubsection{Summary of findings.} The analysis leads to three main conclusions. First, the dominant LP position type is liquidity placement around the current pool price, corresponding to Types 5--6. Second, LPs with consistently positive outcomes in Type 5 positions tend to choose range boundaries such that the current price is located approximately at the midpoint of the interval and do not use the attainment of the range boundaries as a relocation signal. Instead, position closure appears to be driven by a target level of capital return, which on average is reached after the price crosses about \(8\%\) of the selected range in the upward direction. Third, most unique users across the analyzed win-position types (Types 3, 5, and 7) prefer relatively wide liquidity ranges covering price intervals of approximately from 100 to 1000 USDC, whereas the majority of transactions and the largest share of TVL are concentrated in narrow ranges of up to 10 USDC, especially for Type 5 positions. The remaining position types may also provide useful insight, but their interpretation is comparatively more straightforward.



% ... 

% ...


\section{Conclusion}
\label{section:Conclusions}

This paper investigated the economic outcomes of liquidity provision in CLMM pools using historical transaction-level data from the Base network for WETH/USD pools on Uniswap, Aerodrome, PancakeSwap, and SushiSwap. The proposed framework allowed the reconstruction of LP PnL from on-chain events, the classification of LPs by strategy outcome, the introduction of the win-score metric, and the analysis of position types as structural components of LP performance. The results show that, on average, only about one in six LPs avoids losses in the analyzed pools. This suggests that pool rewards alone are unlikely to explain sustained liquidity provision and that other drivers, such as external hedging, portfolio-level considerations, or cross-pool strategies, likely play an important role. In this context, the identification of a distinct class of multi-LPs, i.e., users simultaneously providing liquidity across several pools and protocols, is itself an important result, as it points to the presence of more complex strategies than isolated single-pool participation.

At the position level, the dominant configuration is liquidity placement around the current pool price, and successful LPs typically choose range boundaries such that the current price is located close to the center of the selected interval. At the same time, evidence suggests that classical \(\tau\)-reset strategies are not the predominant form of observed behavior. Successful LPs often close positions before the price reaches the range boundaries, and for the key in-range profitable position type the weighted delta metric is only about \(8\%\). This indicates that position closure is more likely driven by a target level of capital profitability than by full price traversal of the selected range, making the observed behavior closer to \(\tau+\eta\)-type strategies than to standard \(\tau\)-reset logic. Transaction and relocation costs are not modeled separately in the PnL calculation, since they are assumed to be implicitly embedded in the volume of reallocated capital observed in the data. Overall, the results show that standalone liquidity provision is profitable only for a minority of LPs and leave open an important question for future research about the broader economic motives sustaining LP participation despite weak standalone profitability at the pool level.

\section{Acknowledgments}
The authors express their gratitude to the Vega Institute Foundation for organizing and maintaining a research group devoted to optimal control in decentralized finance and to the members of this research group for active and useful discussions. We express special thanks to the VLG Digital team for help with onchain data gathering and insightful research ideas. 


\newpage
\appendix                 
\setcounter{section}{0}   
\renewcommand\thesection{Appendix~\Alph{section}} 

\section{Win-score metric methodology.}\label{app:App11}

To identify \emph{consistently successful} LPs, we introduce the \emph{win-score} metric. For the set of reconstructed position-level PnL values of \(LP_i^c\), ordered by position closing time, we will use the notation
\[
PnL_{\mathrm{Pos}}
=
\left\{
PnL_{j_{M_\alpha}}
\right\}
\]
We then split this set into non-negative and negative subsets, denoted respectively by
\[
\left\{PnL_{\mathrm{Pos}}\right\}^{+}
\quad \text{and} \quad
\left\{PnL_{\mathrm{Pos}}\right\}^{-}.
\]
Positions with non-negative PnL are referred to as \emph{win} positions, while positions with negative PnL are referred to as \emph{loss} positions. To evaluate not only the terminal result, but also the temporal profile of realized performance, we define the cumulative realized PnL path of \(LP_i^c\) as
\[
PnL_{\mathrm{cum}}(t)
=
\sum_{j=1}^{k}
\sum_{\alpha=1}^{p^{\,j}}
PnL_{j_{M_\alpha}}
\cdot
\mathbf{1}\!\left\{
t_{j_{M_\alpha}}\le t
\right\},
\]
where \(t_{j_{M_\alpha}}\) denotes the closing time of the \(\alpha\)-th reconstructed position in range \(r_j\). We then define the positive and negative parts of the cumulative PnL path:
\[
PnL_{\mathrm{cum}}^{\,+}(t)
=
\max\!\left(PnL_{\mathrm{cum}}(t),\,0\right),
\qquad
PnL_{\mathrm{cum}}^{\,-}(t)
=
\min\!\left(PnL_{\mathrm{cum}}(t),\,0\right).
\]

Let
\[
P_{\max}^{\,+}
=
\max_t PnL_{\mathrm{cum}}^{\,+}(t)\ge 0,
\qquad
P_{\min}^{\,-}
=
\min_t PnL_{\mathrm{cum}}^{\,-}(t)\le 0.
\]
Using these extrema, we introduce the normalization factor
\[
S
=
\max\!\left(
P_{\max}^{\,+},
-P_{\min}^{\,-}
\right).
\]
The normalized positive and negative components of the cumulative PnL path are then defined as
\[
C_{+}(t)
=
\frac{
PnL_{\mathrm{cum}}^{\,+}(t)
}{
S
},
\qquad
C_{-}(t)
=
-\frac{
PnL_{\mathrm{cum}}^{\,-}(t)
}{
S
}.
\]
Let \(t_{\mathrm{start}}\) and \(t_{\mathrm{end}}\) denote the beginning and the end of the historical observation period for the LP strategy within the selected sample. Their corresponding areas over this period are
\[
A_{+}
=
\int_{t_{\mathrm{start}}}^{t_{\mathrm{end}}} C_{+}(t)\,dt,
\qquad
A_{-}
=
\int_{t_{\mathrm{start}}}^{t_{\mathrm{end}}} C_{-}(t)\,dt.
\]
Finally, the \emph{win-score} of \(LP_i^c\) is defined as
\[
\mathrm{\omega}
=
\frac{
A_{+}
}{
A_{+}+A_{-}
}.
\]
If \(S=0\), the win-score is set to \(0.5\) as a neutral value.



\clearpage
\begin{sidewaystable}[p]
\section{All Metrics for Win-Position Types}\label{app:App22}
\centering
\label{tab:appendix_b_win_types}
\scriptsize
\setlength{\tabcolsep}{3pt}
\renewcommand{\arraystretch}{1.1}

\resizebox{\textheight}{!}{%
\begin{tabular}{@{}cl*{17}{c}@{}}
\toprule
\multirow{2}{*}{\makecell{Pos.\\type}} &
\multirow{2}{*}{\makecell{Range width\\(USDC)}} &
\multirow{2}{*}{LP} &
\multirow{2}{*}{Trans.} &
\multirow{2}{*}{TVL} &
\multirow{2}{*}{\makecell{Pos. size\\(USDC)}} &
\multirow{2}{*}{\makecell{Duration,\\sec}} &
\multicolumn{2}{c}{\makecell{Range width}} &
\multirow{2}{*}{\makecell{$\Delta$, \%\\(USDC)}} &
\multirow{2}{*}{\makecell{$\Delta_{\mathrm{start}}$}} &
\multirow{2}{*}{$m_a$} &
\multirow{2}{*}{$m_b$} &
\multirow{2}{*}{$d_a$} &
\multirow{2}{*}{$d_b$} &
\multirow{2}{*}{$rel_a$} &
\multirow{2}{*}{$rel_b$} &
\multirow{2}{*}{$R_{\mathrm{cap}}$} &
\multirow{2}{*}{$R_{\mathrm{price}}$} \\
\cmidrule(lr){8-9}
& & & & & & & ticks & USDC & & & & & & & & & & \\
\midrule

\multirow{5}{*}{1}
& $(0,10)$        & 4.4\%  & 2.5\%  & 0.2\%   & 4 183   & 470       & 12     & 3      & --            & --    & -5.0\%  & -5.1\%  & -4.9\%  & -5.0\%  & 0.1\%  & 0.1\%  & 0.1\%  & 0.1\% \\
& $[10,100)$      & 41.2\% & 47.0\% & 3.8\%   & 5 563   & 229 193   & 116    & 34     & --            & --    & -21.4\% & -22.4\% & -20.3\% & -21.3\% & 1.1\%  & 1.1\%  & 2.1\%  & 2.0\% \\
& $[10^2,10^3)$   & 32.4\% & 35.1\% & 93.3\%  & 183 736 & 176 572   & 1 508  & 370    & --            & --    & -5.7\%  & -18.8\% & -4.6\%  & -17.9\% & 1.1\%  & 0.9\%  & 1.2\%  & 1.2\% \\
& $[10^3,10^4)$   & 20.3\% & 11.9\% & 2.8\%   & 16 265  & 435 090   & 6 092  & 3 930  & --            & --    & -30.5\% & -61.7\% & -27.1\% & -59.6\% & 3.4\%  & 2.0\%  & 6.8\%  & 6.8\% \\
& $>10^4$         & 1.6\%  & 3.4\%  & 0.0\%   & 10      & 224 838   & 2 676  & $>10^4$ & --           & --    & -73.3\% & -99.9\% & -99.9\% & -99.9\% & 0.0\%  & 0.0\%  & 0.4\%  & 0.5\% \\
\midrule

\multirow{5}{*}{3}
& $(0,10)$        & 8.9\%  & 7.5\%  & 5.5\%   & 99 328  & 1 807     & 22     & 6      & 50\% (3)      & --    & -0.1\%  & -0.3\%  & 0.1\%   & -0.1\%  & 0.2\%  & 0.2\%  & 0.1\%  & 0.2\% \\
& $[10,100)$      & 17.1\% & 50.4\% & 87.6\%  & 233 281 & 76 020    & 120    & 28     & 31\% (9)      & --    & -0.1\%  & -1.3\%  & 0.4\%   & -0.8\%  & 0.5\%  & 0.5\%  & 0.4\%  & 0.5\% \\
& $[10^2,10^3)$   & 59.3\% & 33.9\% & 6.5\%   & 25 661  & 722 239   & 1 747  & 474    & 39\% (175)    & --    & -0.8\%  & -16.3\% & 7.3\%   & -9.9\%  & 8.0\%  & 6.4\%  & 6.1\%  & 8.2\% \\
& $[10^3,10^4)$   & 13.8\% & 7.9\%  & 0.4\%   & 7 218   & 2 063 699 & 6 880  & 2 240  & 28\% (435)    & --    & -0.3\%  & -46.2\% & 25.9\%  & -33.6\% & 26.2\% & 12.6\% & 17.0\% & 26.3\% \\
& $>10^4$         & 0.8\%  & 0.4\%  & 0.001\% & 202     & 14 580    & 79 766 & $>10^4$ & 0.0005\% (25) & --   & -0.03\% & -99.97\%& 1.4\%   & -99.97\%& 1.4\%  & 0.0\%  & 1.4\%  & 1.4\% \\
\midrule

\multirow{5}{*}{5}
& $(0,10)$        & 0.4\%  & 75.0\% & 58.5\%  & 178 814 & 70 524    & 30     & 8      & 8\% (1)       & 49.5\% & 0.1\%   & -0.2\%  & 0.2\%   & -0.1\%  & 0.02\% & 0.0\%  & 0.01\% & 0.0\% \\
& $[10,100)$      & 7.8\%  & 23.0\% & 41.1\%  & 411 050 & 15 963    & 102    & 27     & 9\% (2)       & 43.0\% & 0.4\%   & -0.6\%  & 0.5\%   & -0.5\%  & 0.1\%  & 0.1\%  & 0.05\% & 0.1\% \\
& $[10^2,10^3)$   & 67.9\% & 1.4\%  & 0.3\%   & 54 801  & 331 258   & 1 816  & 448    & 21\% (83)     & 44.6\% & 9.1\%   & -9.0\%  & 12.8\%  & -6.0\%  & 3.7\%  & 3.0\%  & 1.5\%  & 3.4\% \\
& $[10^3,10^4)$   & 15.5\% & 0.6\%  & 0.03\%  & 12 381  & 865 781   & 7 342  & 1 692  & 13\% (201)    & 46.0\% & 404.9\% & -25.2\% & 450.8\% & -19.5\% & 20.0\% & 5.7\%  & 3.2\%  & 8.0\% \\
& $>10^4$         & 8.3\%  & 0.1\%  & 0.001\% & 528     & 2 108 464 & $>10^6$ & $>10^4$ & 0.006\% (276) & 0.1\% & $>10^4$\% & -99.9\% & $>10^4$\% & -99.9\% & $>10^4$\% & 0.01\% & 6.1\% & 12.7\% \\
\midrule

\multirow{5}{*}{7}
& $(0,10)$        & 4.8\%  & 4.5\%  & 5.9\%   & 37 058  & 886 633   & 24     & 7      & 42\% (3)      & 58.1\% & 0.1\%   & -0.1\%  & 6.7\%   & 6.5\%   & 6.6\%  & 6.5\%  & 1.84\% & 6.6\% \\
& $[10,100)$      & 17.3\% & 26.2\% & 75.1\%  & 81 951  & 204 599   & 171    & 44     & 37\% (18)     & 62.6\% & 1.0\%   & -0.7\%  & 3.6\%   & 1.8\%   & 2.6\%  & 2.5\%  & 0.61\% & 2.5\% \\
& $[10^2,10^3)$   & 75.7\% & 67.8\% & 18.3\%  & 7 713   & 633 072   & 1 267  & 303    & 46\% (134)    & 53.9\% & 7.9\%   & -5.1\%  & 18.1\%  & 3.4\%   & 10.2\% & 8.6\%  & 1.9\%  & 9.3\% \\
& $[10^3,10^4)$   & 2.3\%  & 1.5\%  & 0.8\%   & 14 175  & 1 962 903 & 6 419  & 1 196  & 33\% (377)    & 66.8\% & 1413\%  & -12.9\% & 1577\%  & 3.0\%   & 163.2\%& 15.9\% & 4.5\%  & 20.1\% \\
& $>10^4$         & 0.0\%  & 0.0\%  & 0.0\%   & --      & --        & --     & --     & --            & --    & --      & --      & --      & --      & --     & --     & --     & -- \\
\midrule

\multirow{5}{*}{9}
& $(0,10)$        & 23.0\% & 19.7\% & 39.3\%  & 30 194  & 283 570   & 16     & 4      & 100\% (4)     & --    & -0.2\%  & -0.4\%  & 6.6\%   & 6.4\%   & 6.8\%  & 6.8\%  & 1.32\% & 6.8\% \\
& $[10,100)$      & 33.5\% & 48.9\% & 17.5\%  & 5 441   & 548 788   & 133    & 31     & 100\% (31)    & --    & -1.5\%  & -2.8\%  & 4.5\%   & 3.1\%   & 6.1\%  & 6.0\%  & 3.00\% & 6.3\% \\
& $[10^2,10^3)$   & 41.9\% & 30.1\% & 42.8\%  & 21 608  & 769 783   & 976    & 243    & 100\% (243)   & --    & -0.7\%  & -9.7\%  & 14.1\%  & 3.2\%   & 14.8\% & 12.9\% & 6.9\%  & 15.1\% \\
& $[10^3,10^4)$   & 1.6\%  & 1.3\%  & 0.4\%   & 4 649   & 3 032 835 & 5 330  & 1 061  & 100\% (1061)  & --    & -1.0\%  & -41.6\% & 73.0\%  & 1.5\%   & 73.9\% & 43.2\% & 33.7\% & 74.3\% \\
& $>10^4$         & 0.0\%  & 0.0\%  & 0.0\%   & --      & --        & --     & --     & --            & --    & --      & --      & --      & --      & --     & --     & --     & -- \\
\bottomrule
\end{tabular}%
}
\end{sidewaystable}

\end{document}